\documentclass[aps, twocolumn, letterpaper]{revtex4}

\usepackage{amsmath}
\usepackage{amssymb}
\usepackage{mathrsfs}
\usepackage{xspace}
\usepackage{graphicx}
\usepackage{braket}

\usepackage{ifpdf}
\ifpdf
\fi

\newcommand{\eq}[1]{(\ref{#1})}
\newcommand{\Eq}[1]{Eq.~(\ref{#1})}
\newcommand{\Eqs}[1]{Eqs.~(\ref{#1})}

\newcommand{\Sec}[1]{Sec.~\ref{#1}}
\newcommand{\Ref}[1]{Ref.~\cite{#1}}
\newcommand{\Refs}[1]{Refs.~\cite{#1}}

\newcommand{\eg}{{e.g.,\/}\xspace}				
\newcommand{\ie}{{i.e.,\/}\xspace}				

\newcommand{\pd}{\partial}
\newcommand{\del}{\vec{\nabla}}

\newcommand{\mc}[1]{\mathcal{#1}}
\newcommand{\mcc}[1]{\mathfrak{#1}}

\renewcommand{\vec}[1]{{\boldsymbol{\rm #1}}}

\begin{document}

\title{On the correspondence between quantum and classical variational principles}

\author{D.~E. Ruiz and I.~Y. Dodin}
\affiliation{Department of Astrophysical Sciences, Princeton University, Princeton, New Jersey 08544, USA}

\begin{abstract}
Classical variational principles can be deduced from quantum variational principles via formal reparameterization of the latter. It is shown that such reparameterization is possible without invoking any assumptions other than classicality and without appealing to dynamical equations. As examples, first principle variational formulations of classical point-particle and cold-fluid motion are derived from their quantum counterparts for Schr\"{o}dinger, Pauli, and Klein-Gordon particles.
\end{abstract}

\maketitle

\bibliographystyle{full}

\section{Introduction}

It is commonly known that variational methods are a powerful tool for studying the dynamics of various physical systems, \eg quantum molecular dynamics \cite{ref:feldmeier00, tex:grabowski14}, fluid mechanics \cite{ref:seliger68, ref:morrison98}, classical motion of single particles and collective processes in plasmas \cite{ref:brizard09, book:friedberg}, and wave propagation in both linear and nonlinear media \cite{book:whitham, book:tracy, book:joannopoulos, my:amc}. Variational formulations are advantageous as they lead to dynamical equations in a manifestly conservative form derived from a single scalar function, a Lagrangian (or Lagrangian density). The fact that these equations can be approximated robustly and self-consistently by approximating just one function makes the method particularly attractive for reduced calculations \cite{ref:brizard09, my:itervar, my:lens}. However, so far, exact Lagrangians have been obtained largely heuristically or \textit{ad~hoc} \cite{foot:adhoc}. Such approaches tend to obscure the physical meaning of the results and limit their applicability, while regular ways to deduce Lagrangians rigorously from first principles are yet to be found.

Here we show that classical variational principles (VPs) can be deduced from quantum VPs, which are well known, via formal reparameterization of the latter. Such reparameterization is possible without appealing to dynamical equations and without invoking any assumptions other than classicality. This distinguishes our theory from the existing variational formulations of the quantum-classical correspondence, which are more restrictive \cite{foot:other}. Also as a complement to those formulations, we consider both single-particle \textit{and} fluid VPs and rigorously explain how they are connected through quantum VPs. Classical-fluid Lagrangians flow as the semiclassical limit of the fundamental quantum Lagrangian (FQL), \Eq{eq:gl}, and the point-particle Lagrangians are then yielded as corollaries for narrow \textit{but otherwise arbitrary} wave packets. This approach enables the first principle classical Lagrangian description of most general, relativistic vector particles (\eg a Dirac electron) and, similarly, the geometrical optics (GO) description of any vector waves, as we discuss in a companion paper \cite{foot:myqdirac}. The present paper is intended as an introduction to such calculations. Thus, below, we primarily focus on systematization of VPs for commonly known systems, including Schr\"{o}dinger, Pauli, and Klein-Gordon particles. 

We explicitly show how to deduce classical Lagrangians for these particles and the corresponding fluids from the FQL. In particular, we show that the expression for the Bohm quantum potential \cite{ref:bohm52, ref:bohm52b}, the so-called Weizs\"{a}cker correction \cite{ref:weizsacker35}, the Madelung equations \cite{ref:madelung27}, and the Chen-Sudan Lagrangian density \cite{ref:chen93} all emerge naturally within our unifying theory as special cases. We also obtain an alternative, manifestly Lagrangian representation of Takabayasi equations for a Pauli particle. In contrast to the original theory \cite{ref:takabayasi55}, our model yields the dynamics of the spin vector $\vec{S}$ directly from the VP and employs two, rather than three, equations; hence, $S = 1/2$ is ensured irrespective of initial conditions. 

The paper is organized as follows. In \Sec{sec:notation}, we define the basic notation. In \Sec{sec:basic}, we introduce the general formulas. In \Sec{sec:schrodinger}, we discuss a Schr\"{o}dinger particle and the Madelung equations. In \Sec{sec:pauli}, we discuss a Pauli particle and the variational formulation of Takabayasi equations. In \Sec{sec:kleingordon}, we discuss a Klein-Gordon particle and rederive the Chen-Sudan fluid Lagrangian density. In \Sec{sec:conclusion}, we summarize our main results and outline future applications of the proposed theory.

\section{Notation}
\label{sec:notation}

The following notation is used throughout the paper. The symbol ``$\doteq$'' denotes definitions. We use natural units, so the speed of light equals one ($c = 1$), and so is the Planck constant ($\hbar = 1$). The Minkowski metric is adopted with signature $(-, +, +, +)$, so, in particular, $\mathrm{d}^4x \equiv \mathrm{d}t\,\mathrm{d}^3x$. Generalizations to curved metrics are straightforward to apply \cite{my:wkin}. Greek indexes span from $0$ to $3$ and refer to spacetime coordinates, $x^\mu$, with $x^0$ corresponding to the time variable, $t$; in particular, $\pd_\mu \equiv \pd/\pd x^\mu$. Latin indexes span from $1$ to $3$ and denote the spatial variables, $x^i$ (except where specified otherwise); in particular, $\pd_i \equiv \pd/\pd x^i$. Summation over repeated indexes is assumed. Also, in the Euler-Lagrange equations (ELEs), the notation ``$\delta a: $'' denotes, as usual, that the corresponding equation was obtained by extremizing the action integral with respect to $a$.

\section{Basic equations}
\label{sec:basic}

\subsection{Fundamental quantum Lagrangian}

The dynamics of a quantum particle is governed by the least action principle, 
\begin{gather}\label{eq:lap}
\delta \Lambda = 0,
\end{gather}
where $\Lambda$ is the action integral,
\begin{gather}\label{general_action}
\Lambda = \int \mcc{L}\,\mathrm{d}^4x,
\end{gather}
and $\mcc{L}$ is the Lagrangian density. Like for any other nondissipative linear wave \cite{my:wkin}, the Lagrangian density of a quantum particle can be expressed as \cite{foot:df},
\begin{gather}\label{eq:gl}
\mcc{L} = \frac{i}{2}\,[\psi^\dag (\pd_t \psi) - (\pd_t \psi^\dag)\psi] - \psi^\dag \hat{H} \psi,
\end{gather}
which is what we term (the density of) the FQL. Here $\hat{H}$ is some Hermitian operator called Hamiltonian, $\psi$ is a complex vector field (``state function''), and $\psi^\dag$ is its adjoint. Treating $\psi$ and $\psi^\dag$ as independent functions \cite{my:wkin}, one obtains from \Eq{eq:lap} two mutually adjoint ELEs,
\begin{align}
\delta \psi^\dag & : \quad i \pd_t \psi = \hat{H} \psi, \\
\delta \psi & : \quad -i \pd_t \psi^\dag = (\hat{H}\psi)^\dag .
\end{align}

It is also possible to derive other, different equations from \Eqs{eq:gl} if $\mcc{L}$ is reparameterized, \ie if other independent functions are chosen instead of $\psi$ and $\psi^\dag$. This can be done in general, using the fact that the Hamiltonian always can be expressed as $\hat{H} = H(t, \hat{\vec{x}}, \hat{\vec{p}})$ \cite{my:wkin} (here $H$ is some matrix function, $\hat{\vec{x}} = \vec{x}$ is the position operator, $\hat{\vec{p}} = -i \del$ is the momentum operator, and the standard coordinate representation is assumed), so
\begin{gather}\label{eq:gl2}
\mcc{L} = \frac{i}{2}\,[\psi^\dag (\pd_t \psi) - (\pd_t \psi^\dag)\psi] - \psi^\dag H(t, \vec{x}, -i \del) \psi.
\end{gather}
Specifically, the classical reparameterization of this Lagrangian density is performed as follows.

\subsection{Single-particle Lagrangian}
\label{sec:single}

First, let us consider the simplest classical reparameterization of \Eq{eq:gl2}. For a scalar particle, the state function can be expressed as
\begin{gather}\label{eq:psipar}
\psi = a e^{i\theta},
\end{gather}
where the amplitude $a \equiv \sqrt{\mc{I}}$ and the phase $\theta$ are real scalar functions of $(t, \vec{x})$. Let us assume the semiclassical limit, \ie that $\mc{I}(t, \vec{x})$ is slow compared to $\theta(t, \vec{x})$. Then, \Eq{eq:gl2} gives \cite{foot:pose}
\begin{gather}\label{eq:hayes}
\mcc{L} \approx - \mc{I}[\pd_t \theta + H(t, \vec{x}, \del \theta)],
\end{gather}
which one may recognize as Hayes's representation of the Lagrangian density of a GO wave \cite{ref:hayes73}. Treating $\theta$ and $\mc{I}$ as independent functions, one hence arrives at the following ELEs:
\begin{align}
\delta \theta & : \quad \pd_t \mc{I} + \del \cdot (\mc{I} \vec{v}) = 0, \label{eq:act}\\
\delta \mc{I} & : \quad \pd_t \theta + H(t, \vec{x}, \del \theta) = 0, \label{eq:hj}
\end{align}
where $\vec{v} \doteq \pd_\vec{p} H(t, \vec{x}, \vec{p})$ is the group velocity evaluated at the wave vector $\vec{p}(t, \vec{x}) = \del \theta$. Equation \eq{eq:act} is a continuity equation known as the action conservation theorem (ACT). Equation \eq{eq:hj} is the Hamilton-Jacobi representation of the local dispersion relation,
\begin{gather}
\mc{E} = H(t, \vec{x}, \vec{p}),
\end{gather}
where $\mc{E} \doteq - \pd_t \theta$ is the local frequency.

To the extent that a wave packet is well localized such that it is meaningful to describe its dynamics as the dynamics of the packet's geometrical center, the continuous-wave description can be replaced with a simpler, point-particle model. In this case, one can approximate the action density with a delta function,
\begin{gather}\label{eq:delta}
\mc{I}(t, \vec{x}) = \delta(\vec{x} - \vec{X}(t)),
\end{gather}
so the LD can be replaced with just a point-particle Lagrangian, $L \doteq \int \mcc{L}\,\mathrm{d}^3x$. This is done as follows. Let us express the complete action integral, $\Lambda$, in the form
\begin{multline}
\Lambda = \underbrace{- \int \mc{I}(t, \vec{x})\, \pd_t \theta(t, \vec{x})\,\mathrm{d}^4x}_{\Lambda_1} \\
\underbrace{- \int \mc{I}(t, \vec{x})\,H(t, \vec{x}, \del \theta(t, \vec{x}))\,\mathrm{d}^4x}_{\Lambda_2}.
\end{multline}
Then, substituting \Eq{eq:delta} leads to
\begin{align} 
\Lambda_1  = & - \int \delta(\vec{x} - \vec{X}(t))\,\pd_t \theta(t, \vec{x})\,\mathrm{d}^4x \notag\\
 = & \int \left[\pd_t \delta(\vec{x} - \vec{X}(t))\right] \theta(t, \vec{x})\,\mathrm{d}^4x \notag\\
 = & \int \left[-\del \delta(\vec{x} - \vec{X}(t))\right] \cdot \dot{\vec{X}}(t)\, \theta(t, \vec{x})\,\mathrm{d}^4x \notag\\
 = & \int \left[ - \int \theta(t, \vec{x})\, \del \delta(\vec{x} - \vec{X}(t))\,\mathrm{d}^3x \right] \cdot \dot{\vec{X}}(t)\,\mathrm{d}t \notag\\
 = & \int \left[ \int \delta(\vec{x} - \vec{X}(t)) \del \theta(t, \vec{x})\,\mathrm{d}^3x \right] \cdot \dot{\vec{X}}(t)\,\mathrm{d}t \notag\\
 = & \int \vec{P}(t) \cdot \dot{\vec{X}}(t)\,\mathrm{d}t,
\label{eq:s1del}
\end{align}
where $\vec{P}(t) \doteq \del \theta(t, \vec{X}(t))$. Similarly,
\begin{gather}
\Lambda_2 = - \int H(t, \vec{X}(t), \vec{P}(t))\,\mathrm{d}t.
\end{gather}
Thus, the total action integral is expressed as $\Lambda = \int L\,\mathrm{d}t$. The function $L$ serves as a point-particle Lagrangian and is given by
\begin{gather}\label{eq:LC}
L = \vec{P} \cdot \dot{\vec{X}} -  H(t, \vec{X}, \vec{P}).
\end{gather}

One may recognize this $L$ as the standard classical Lagrangian, where $\vec{X}$ and $\vec{P}$ serve as a canonical coordinate and a canonical momentum, respectively. Treating them as independent variables then leads \cite[Sec.~43]{book:landau1} to ELEs matching Hamilton's equations in the canonical form,
\begin{align}
\delta \vec{P} : & \quad \dot{\vec{X}} = \pd_\vec{P} H(t, \vec{X}, \vec{P}), \label{eq:canX}\\
\delta \vec{X} : & \quad \dot{\vec{P}}=  - \pd_\vec{X} H(t, \vec{X}, \vec{P}). \label{eq:canP}
\end{align}
(Hence, $\mc{E}$ and $\vec{p}$ can be understood as local canonical energy and momentum.) Equations \eq{eq:canX} and \eq{eq:canP} also serve as ray equations for \Eqs{eq:act} and \eq{eq:hj} \cite{my:wkin}. 

It is to be noted that our derivation does not require interpreting $\psi$ as a probability and holds for any $H$. It is also to be noted that our point-particle equations are derived without assuming a specific profile of $\mc{I}(t, \vec{x})$. All we require instead is that this function be narrow enough (yet wide enough compared to the particle wavelength; otherwise, the GO approximation does not apply).

\subsection{Fluid Lagrangian density}
\label{sec:fluid}

Now consider an \textit{ensemble} of particles described by $(\theta_q, \mc{I}_q)$, $q = 1,...\,N$. Assuming the energy of particle interactions (except through average fields included in $H$) is negligible, the ensemble Lagrangian density $\mcc{L}$ equals the sum of single-particle Lagrangians $\mcc{L}_q$ given by \Eq{eq:hayes}. Let us also assume that particles are cold. This means that all $\vec{p}_q$ are about the same, and thus so are all $\mc{E}_q$; \ie 
\begin{gather}
\del \theta_q \approx \del \theta, \quad \pd_t \theta_q \approx \pd_t \theta, 
\end{gather}
where $\theta$ is some average phase. Then, the ensemble is described by the same $\mcc{L}$ as the one given by \Eq{eq:hayes}, except $\mc{I}$ must be replaced with the particle density, 
\begin{gather}
n \doteq \sum_{q = 1}^N \mc{I}_q. 
\end{gather}
In other words, the ensemble's Lagrangian density is given~by
\begin{gather}\label{eq:hayesf}
\mcc{L} = -n[\pd_t \theta + H(t, \vec{x}, \del \theta)].
\end{gather}
The corresponding ELEs are
\begin{align}
\delta \theta & : \quad \pd_t n + \del \cdot (n \vec{v}) = 0, \label{eq:actf}\\
\delta n & : \quad \pd_t \theta + H(t, \vec{x}, \del \theta) = 0, \label{eq:hjf}
\end{align}
and $\vec{v}$ serves as the flow velocity. These equations provide a complete description of the ensemble as a cold fluid, so $\mcc{L}$ given by \Eq{eq:hayesf} can be viewed as the Lagrangian density of a cold classical fluid. (Specific examples will be discussed in the next sections.) It is to be noted, like we already did before, that interpreting $\psi$ as a probability amplitude is not needed within this approach.

Thermal corrections for isentropic fluid also can be introduced readily, at least \textit{ad~hoc}. This is done by adding to $\mcc{L}$ a term $- n U(t, \vec{x}, n)$, which leads to
\begin{gather}\label{eq:hayesf2}
\mcc{L} = -n[\pd_t \theta + H(t, \vec{x}, \del \theta) + U(t, \vec{x}, n)].
\end{gather}
From comparing \Eqs{eq:hayesf} and \eq{eq:hayesf2}, it is seen that $U$ has the meaning of the per-particle internal energy; then $n^2 \pd_n U$ serves as the pressure \cite{foot:pressure}. 

Outside the \textit{ad~hoc} approach, there exists other variational methods to obtain corrections to the cold-fluid variational principle that we derived. In \Ref{Michta:2015ww}, the Thomas-Fermi energy functional is minimized in order to obtain the quantum hydrodynamic equations with thermal gradient corrections included. Also, the Lagrangian density for more general, non-isentropic fluids \cite{ref:seliger68} can be deduced from the FQL. Details will be reported in a separate paper. Below, we will completely ignore thermal effects for clarity.

\section{Schr\"{o}dinger particle}
\label{sec:schrodinger}

Let us specifically discuss a Schr\"{o}dinger particle, \ie a nonrelativistic charged particle governed by
\begin{gather}\label{eq:schH}
\hat{H} = \frac{1}{2m}\,(-i\del - q\vec{A})^2 + q \varphi.
\end{gather}
Here $m$ and $q$ are the particle mass and charge, $\vec{A} = \vec{A}(t, \vec{x})$ is the vector potential, and $\varphi = \varphi(t, \vec{x})$ is the scalar potential. Substitution of \Eq{eq:psipar} leads to the following \textit{exact} Lagrangian density,
\begin{gather}
\mcc{L} = - \mc{I} \left[\pd_t \theta +  \frac{1}{2m}( \del \theta - q \vec{A})^2 +  q \varphi \right]- W_{\rm S}, \label{eq:madL}\\
W_{\rm S} \doteq \frac{(\del a)^2}{2m} = \frac{(\del \mc{I})^2}{8m \mc{I}},
\end{gather}
where $W_{\rm S}$ can be recognized as the Weizs\"{a}cker correction \cite{ref:weizsacker35}. The corresponding ELEs are
\begin{align}
\delta \theta : & \quad \pd_t \mc{I} + \del \cdot (\mc{I} \vec{v}) = 0, \label{eq:actschr}\\
\delta \mc{I} : & \quad \pd_t \theta + \frac{1}{2m}(\del \theta - q \vec{A})^2  + q \varphi + Q = 0, \label{eq:hjschr}
\end{align}
where $\vec{v} \doteq (\del \theta - q \vec{A})/m$, and
\begin{equation}
Q \doteq  -  \frac{\del^2 \sqrt{\mc{I}}}{2m\sqrt{\mc{I}}}
\end{equation}
is recognized as the Bohm quantum potential \cite{ref:bohm52, ref:bohm52b}, which is obtained when varying the $\int W_{\rm S}\, \mathrm{d}^4 x$ term of the action. Equation \eq{eq:actschr} represents the ACT, and \Eq{eq:hjschr} represents a quantum Hamilton-Jacobi equation. Taking the gradient of the latter readily yields
\begin{equation}\label{eq:madelung}
m (\pd_t + \vec{v} \cdot \del) \vec{v}  = q (\vec{E} + \vec{v} \times \vec{B}) - \del Q.
\end{equation}

Equations \eq{eq:actschr} and \eq{eq:madelung} combined together coincide with Madelung equations \cite{ref:madelung27}. Hence, $\mcc{L}$ given by \Eq{eq:madL} can be identified as the Madelung Lagrangian density. The equation for cold classical fluid are obtained by replacing $\mc{I}$ with $n$, as in \Sec{sec:fluid}, and also by neglecting $Q$ in \Eq{eq:madelung} or, equivalently, $W_{\rm S}$ in \Eq{eq:madL} (which constitutes the semiclassical, or GO approximation). Hence,
\begin{gather}
\mcc{L} = - n\left[\pd_t \theta +  \frac{1}{2m}( \del \theta - q \vec{A})^2 + q \varphi \right]
\end{gather}
can be identified as a Lagrangian density of a cold classical fluid, in agreement with \Ref{ref:seliger68}. According to \Sec{sec:single}, the classical point-particle motion is then governed by \Eqs{eq:canX} and \eq{eq:canP} with the Hamiltonian
\begin{gather}
H(t, \vec{X}, \vec{P}) = \frac{1}{2m}\,[\vec{P} - q \vec{A}(t, \vec{X})]^2 + q \varphi(t, \vec{X}).
\end{gather}

\section{Pauli particle}
\label{sec:pauli}

Now let us consider a Pauli particle, \ie a non-relativistic spin-$1/2$ particle governed by \cite{ref:pauli27}
\begin{gather}\label{eq:hpauli}
\hat{H} = \frac{1}{2m} (-i\del - q\vec{A})^2 + q \varphi - \frac{q}{2m}\,\vec{\sigma} \cdot \vec{B},
\end{gather}
where $\vec{\sigma}=(\sigma_x, \sigma_y, \sigma_z)$ are the Pauli matrices,
\begin{align}
\sigma_x =
\begin{pmatrix}
0 & 1 \\
1 & 0 
\end{pmatrix} 
&, &
\sigma_y =
\begin{pmatrix}
0 & -i \\
i & 0 
\end{pmatrix} 
&, &
\sigma_z =
\begin{pmatrix}
1 & 0 \\
0 & -1 
\end{pmatrix},
\end{align}
and $\vec{B} \doteq \del \times \vec{A}$ is magnetic field. [Replacing the Bohr magneton $q/(2m)$ in \Eq{eq:hpauli} with a different constant would not significantly affect the below calculation.] The state function $\psi$ has two complex components; \ie
\begin{gather}\notag
\psi =
\begin{pmatrix}
a_1 e^{i\theta_1} \\
a_2 e^{i\theta_2}
\end{pmatrix},
\end{gather}
where $a_{1, 2}$ and $\theta_{1, 2}$ are real. Thus, it can be cast in the following general form, $\psi = e^{i\theta}\eta(\vartheta, \zeta) \sqrt{\mc{I}}$, with
\begin{gather}\notag
\eta(\vartheta, \zeta) = 
\begin{pmatrix}
e^{-i\vartheta/2} \cos (\zeta/2) \\
e^{i\vartheta/2} \sin (\zeta/2)
\end{pmatrix}.
\end{gather}
Here $\theta\doteq (\theta_1 + \theta_2)/2$, $\vartheta \doteq \theta_2 - \theta_1$, $\mc{I} \doteq \psi^\dag \psi$, $\zeta$ is real and defined via 
\begin{gather}
a_1 = a \cos (\zeta/2), \quad a_2 = a \sin (\zeta/2),
\end{gather}
and $a \doteq \sqrt{\mc{I}}$. Hence, the Lagrangian density can be expressed as
\begin{multline}\label{eq:lpc}
\mcc{L} = \frac{i\mc{I}}{2} \left[\eta^\dag (d_t \eta) - (d_t \eta^\dag) \eta \right] - \mc{I}(\pd_t \theta) - W_{\rm P}\\
- \mc{I} \left[\frac{1}{2m}\,(\del \theta - q\vec{A})^2 + q \varphi + U_{\rm SG}\right],
\end{multline}
where $d_t \doteq \pd_t + \vec{v} \cdot \del$ is the convective derivative associated with velocity field $\vec{v} \doteq (\del \theta -q\vec{A})/m$, and
\begin{gather}
U_{\rm SG} = - \frac{q}{2m}\,\eta^\dag (\vec{\sigma} \cdot \vec{B}) \eta,\\
W_{\rm P} = \frac{(\del a)^2}{2m} + \frac{\mc{I}}{2m}\,(\del \eta^\dag) \cdot (\del \eta),
\end{gather}
where we made use of the fact that $\eta^\dag\eta = 1$. The term $U_{\rm SG}$ is the Stern-Gerlach energy. It can be expressed conveniently as $U_{\rm SG} = -(q/m)\,\vec{S} \cdot \vec{B}$, where
\begin{gather}\label{eq:spin_vector}
\vec{S} \doteq \frac{1}{2}\,\eta^\dag \vec{\sigma} \eta = 
\frac{1}{2}
\begin{pmatrix}
\sin \zeta \, \cos \vartheta \\
\sin \zeta \, \sin \vartheta \\
\cos \zeta 
\end{pmatrix}
\end{gather}
is the spin vector, $S = 1/2$. Expressing also the other terms through the four independent variables, $(\theta, \mc{I}, \vartheta, \zeta)$, one gets
\begin{multline}\notag
\mcc{L} = - \mc{I}\bigg[\pd_t \theta + \frac{1}{2m}\,(\del \theta - q \vec{A})^2 + q\varphi \\
- \frac{1}{2}(d_t \vartheta)\,\cos \zeta - \frac{q}{m}\,\vec{S}(\vartheta, \zeta) \cdot \vec{B}\bigg] - W_{\rm P},
\end{multline}
\begin{gather}\label{eq:wp}
W_{\rm P} = \frac{(\del \mc{I})^2}{8m\mc{I}} + \frac{\mc{I}}{8m}\,[(\del \zeta)^2 + (\del \vartheta)^2].
\end{gather}

Hence, the following four ELEs are yielded. The first one is the ACT,
\begin{gather}
\delta \theta : \quad \pd_t \mc{I} + \del \cdot (\mc{I} \vec{V}) = 0,
\label{eq:actp}
\end{gather}
where the flow velocity is given by $\vec{V} = \vec{v} + \vec{u}$, and
\begin{gather}\notag
\vec{u} 
\doteq - \frac{i}{2m}\, [\eta^\dag (\del \eta) -(\del \eta^\dag) \eta]
= - \frac{1}{2m}\,(\del \vartheta) \cos \zeta.
\end{gather}
The second ELE is a Hamilton-Jacobi equation,
\begin{multline}\label{eq:momp}
\delta \mc{I} : \quad \pd_t \theta + \frac{1}{2m}\,(\del \theta - q \vec{A})^2  + q\varphi + Q - \frac{q}{m}\,\vec{S} \cdot \vec{B} \\
 = \frac{1}{2}\,(d_t \vartheta) \cos \zeta - \frac{1}{8m}\,[(\del \zeta)^2 + (\del \vartheta)^2].
\end{multline}
Finally, the remaining two ELEs are
\begin{multline}
\delta \vartheta : \quad 
\pd_t (\mc{I} \cos \zeta)  + \del \cdot (\mc{I} \vec{v} \cos \zeta)\\
= \frac{1}{2m}\,\del \cdot (\mc{I} \del \vartheta) + \frac{2q \mc{I}}{m}\,(\pd_\vartheta \vec{S}) \cdot \vec{B},
\label{eq:dvartheta}
\end{multline}
and
\begin{gather}
\delta \zeta : \quad \mc{I}\,(d_t \vartheta) \sin\zeta =
\frac{1}{2m}\,\del \cdot (\mc{I} \del \zeta) + \frac{2q \mc{I}}{m}\,(\pd_\zeta \vec{S}) \cdot \vec{B}.
\label{eq:dzeta}
\end{gather}

As shown in appendix, \Eqs{eq:momp}-\eq{eq:dzeta} represent an alternative form of Takabayasi equations \cite{ref:takabayasi55},
\begin{multline}
m (\pd_t + \vec{V} \cdot \del) \vec{V} = q(\vec{E} + \vec{V} \times \vec{B}) - \del Q \\
+ \frac{q}{m}\, S^i(\del B_i) - \frac{1}{m\mc{I}}\, \pd_j \left[\mc{I} (\del S_i) (\pd^j S^i) \right],
\label{eq:takamom}
\end{multline}
\begin{gather}
\pd_t \vec{S} + (\vec{V} \cdot \del) \vec{S} = \frac{q}{m}\,\vec{S} \times \vec{B} 
+ \frac{1}{m\mc{I}}\, \vec{S} \times \pd_i(\mc{I}\, \pd^i \vec{S})
\label{eq:takaspin}
\end{gather}
(which were also studied recently in \Refs{ref:stefan11, ref:brodin11b, ref:zamanian10b, ref:brodin07, ref:marklund07, foot:lingam}). However, our equations have a number of advantages: (i) they have a manifestly Lagrangian form; (ii) the spin dynamics is described by two rather than three equations; (iii) $S = 1/2$ is ensured irrespective of initial conditions.

The Lagrangian density for cold fluid is obtained by replacing $\mc{I}$ with $n$, as in \Sec{sec:fluid}, and by assuming that the real angles, which describe the particle spin, are about the same; \ie $\vartheta_q \approx \vartheta$ and $\zeta_q \approx \zeta$, where $\vartheta$ and $\zeta$ are the corresponding average angles. Hence, 
\begin{multline}
\mcc{L} = - n \bigg[\pd_t \theta + \frac{1}{2m}\,(\del \theta - q \vec{A})^2 + q\varphi \\
 - \frac{1}{2}\,(d_t \vartheta) \cos \zeta - \frac{q}{m}\,\vec{S}(\vartheta, \zeta) \cdot \vec{B}\bigg]-W_{\rm P} 
\end{multline}
can be identified as a Lagrangian density of a cold fluid comprised of Pauli particles. The point-particle Lagrangian is derived much like in \Sec{sec:single}. We neglect the Bohm quantum potential and also assume that $\vartheta$ and $\zeta$ are approximately homogeneous in the regions in which $\mc{I}(t,\vec{x})$ is nonzero. Specifically, one obtains
\begin{gather}
L = \vec{P} \cdot \dot{\vec{X}}+ \frac{1}{2}\,\dot{\vartheta} \cos \zeta - H(t, \vec{X}, \vec{P}, \vartheta, \zeta),
\end{gather}
where the Hamiltonian $H$ is given by
\begin{multline}
H(t, \vec{X}, \vec{P}, \vartheta, \zeta) = 
\frac{1}{2m}\,(\vec{P} - q \vec{A}(t, \vec{X}))^2 \\
+ q\varphi(t, \vec{X}) - \frac{q}{m}\,\vec{S}(\vartheta, \zeta) \cdot \vec{B}(t, \vec{X}).
\end{multline}
The corresponding ELEs are
\begin{align}
\delta \vec{P}   : & \quad \dot{\vec{X}} = \pd_\vec{P} H(t, \vec{X}, \vec{P}, \vartheta, \zeta),\\
\delta \vec{X}   : & \quad \dot{\vec{P}}=  - \pd_\vec{X} H(t, \vec{X}, \vec{P}, \vartheta, \zeta), \\
\delta \vartheta : & \quad \dot{\zeta}\,\sin \zeta = -(2q/m)\,(\pd_\vartheta \vec{S}) \cdot \vec{B}, \\
\delta \zeta     : & \quad \dot{\vartheta}\,\sin \zeta = + (2q/m)\,(\pd_\zeta \vec{S}) \cdot \vec{B}.
\end{align}
These equations also yield the following equation for~$\vec{S}$,
\begin{equation}
\dot{\vec{S}} = \frac{q}{m}\, \vec{S} \times \vec{B}(t, \vec{X}),
\end{equation}
as can be checked by direct substitution. Also, even without such calculation, the fact that $\dot{\vec{S}}$ is perpendicular to $\vec{S}$ is anticipated from the definition of the spin vector, \Eq{eq:spin_vector}, according to which $S$ must remain constant.

Notice that, within our approach, the spin precession equation flows directly from the VP. This distinguishes our approach from other calculations (\eg in \Refs{ref:derbenev73, arX:heinemann96}), where the equation for $\vec{S}$ is derived from a postulated Poisson bracket for the classical spin variables. We also notice that our Lagrangian can be represented in terms of complex variables, in a form akin to but different from those in \Refs{ref:barut84, ref:barut85, ref:barut90, ref:barut93}. This complex representation and also a detailed comparison with related point-particle models of spin electrons is discussed in a companion paper \cite{foot:myqdirac}, where our results are extended to relativistic particles.

\section{Klein-Gordon particle}
\label{sec:kleingordon}

It is instructive to discuss also Lagrangian densities of other types. For example, let us consider \cite{foot:fv}
\begin{equation}\label{eq:kglagr}
\mcc{L} = -\frac{1}{2m}\,[(D^\mu\psi)^* (D_\mu \psi) + m^2 \psi^* \psi],
\end{equation}
which corresponds to a Klein-Gordon particle \cite{ref:klein27, ref:gordon26}, \ie a relativistic spinless particle governed by
\begin{equation}\label{eq:kgeq}
(D_\mu D^\mu + m^2) \psi = 0.
\end{equation}
Here $D_\mu \doteq - i\pd_\mu - q A_\mu$, $A_\mu = A_\mu(x^\nu)$ is an electromagnetic four-vector potential (so $\varphi = - A_0$), and $\psi$ is complex scalar function. [The factor $(2m)^{-1}$ in \Eq{eq:kglagr} is added to ensure that $\psi$ is normalized such that $\psi^* \psi$ relates to the action density most transparently; see below.] Let us represent the state function as
\begin{equation}
\psi = e^{i\theta}\sqrt{\mc{I}_0}
\end{equation}
and adopt the semiclassical limit, like in \Sec{sec:basic}. Then,
\begin{gather}\label{eq:kgsemi}
\mcc{L} \approx \frac{\mc{I}_0}{2m}\,[(\pd_t \theta + q \varphi)^2 - (\del \theta -q\vec{A})^2 - m^2].
\end{gather}
This has the form of Whitham's Lagrangian density \cite{my:amc},
\begin{gather}
\mcc{L} = \mcc{D}(
\underbrace{-\pd_t \theta\vphantom{\del \theta}}_{\mc{E}\vphantom{\vec{p}}}, 
\underbrace{\del \theta\vphantom{-\pd_t \theta}}_{\vec{p}\vphantom{\mc{E}}}; t, \vec{x}) \mc{I}_0,
\end{gather}
in which
\begin{gather}
\mc{I} \doteq \mc{I}_0\,\pd_\mc{E} \mcc{D}(\mc{E}, \vec{p}; t, \vec{x}) = \gamma \mc{I}_0
\end{gather}
serves as the action density, and $\gamma \doteq (\mc{E} - q \varphi)/m$. This yields the ACT and a Hamilton-Jacobi equation, 
\begin{align}
\delta \theta : & \quad \pd_t \mc{I} + \del \cdot (\mc{I}\vec{v}) = 0,\label{eq:kgact1} \\
\delta \mc{I} : & \quad (\pd_t \theta + q \varphi)^2 = (\del \theta - q\vec{A})^2 + m^2, \label{eq:kghj}
\end{align}
where the flow velocity $\vec{v}$ is given by
\begin{gather}
\vec{v} \doteq \frac{\del \theta - q\vec{A}}{m\gamma}.
\end{gather}
Hence one may recognize $\gamma$ as the Lorentz factor [because $\mc{E}$ serves as the canonical energy (\Sec{sec:basic}) and, from \Eq{eq:kghj}, it is seen that $\mc{E} - q \varphi$ is the kinetic energy], so $\mc{I}_0$ is recognized as the proper action density.

From \Eq{eq:kghj}, one gets
\begin{gather}
- \pd_t \theta = \pm \sqrt{(\del \theta - q \vec{A})^2  + m^2} + q \varphi.
\end{gather}
To ensure that the previously adopted semiclassical approximation is satisfied, one of the two solutions must be excluded. We will consider positive-energy particles, \ie assume the plus sign. Then the wave is of a scalar type, and, as it is well known \cite{ref:hayes73}, Whitham's Lagrangian density can be rewritten in the Hayes's form \eq{eq:hayes}; namely,
\begin{gather}\label{eq:kgL}
\mcc{L} \approx - \mc{I} \big\{\pd_t \theta + \sqrt{[\vec{p} - q \vec{A}(t, \vec{x})]^2  + m^2} + q \varphi(t, \vec{x})\big\}.
\end{gather}

As in the previous sections, the Lagrangian density of a cold (yet now relativistic) classical fluid is obtained by replacing $\mc{I}$ with $n$. The corresponding ELEs coincide with \Eqs{eq:actf} and \eq{eq:hjf}, and taking the gradient of the latter leads to the well known relativistic momentum equation,
\begin{gather}
[\pd_t + (\vec{v} \cdot \del)] (m\gamma \vec{v})= q(\vec{E} + \vec{v} \times \vec{B}).
\end{gather}
Also, the classical point-particle motion is automatically seen to be governed by \Eqs{eq:canX} and \eq{eq:canP} with
\begin{gather}\label{eq:kgspH}
H(t, \vec{X}, \vec{P}) = \sqrt{[\vec{P} - q \vec{A}(t, \vec{X})]^2  + m^2} + q \varphi(t, \vec{X}).
\end{gather}

Finally, let us discuss the interactions of particles with \textit{self-consistent} fields. The Lagrangian density $\mcc{L}_\Sigma$ of such interactions is obtained, as usual, by adding to $\mcc{L}$ the Lagrangian density of the vacuum field \cite[Sec.~11.5]{book:goldstein},
\begin{gather}\notag
\mcc{L}_{\rm EM} = \frac{1}{8\pi}\,(- \del \varphi - \pd_t \vec{A})^2 - \frac{|\del \times \vec{A}|^2}{8\pi} \equiv \frac{E^2 - B^2}{8\pi},
\end{gather}
where $\vec{E}$ and $\vec{B}$ are electric and magnetic fields. This gives
\begin{gather}\label{eq:LSigma}
\mcc{L}_\Sigma = \mcc{L}_{\rm EM} - \sum_s n_s [\pd_t \theta_s + H_s(t, \vec{x}, \del \theta_s)],
\end{gather}
where the sum is taken over all species. In particular, substituting here the Hamiltonian \eq{eq:kgspH} automatically yields the well-known Chen-Sudan Lagrangian density of cold relativistic plasma \cite{ref:chen93}. It is to be noted that, in the form \eq{eq:LSigma}, the obtained $\mcc{L}_\Sigma$ can be treated as the Lagrangian density of warm plasma too, if particles with different energies are treated as different species; for example, see \Ref{my:itervar}.

\section{Discussion}
\label{sec:conclusion}

We showed that classical VPs can be deduced from quantum VPs, which are well known, via formal reparameterization of the latter. Such reparameterization is possible without appealing to dynamical equations and without invoking any assumptions other than classicality. Classical-fluid Lagrangians flow as the semiclassical limit of the FQL, and the point-particle Lagrangians are then yielded as corollaries for narrow but otherwise arbitrary wave packets. We explicitly performed these calculations for several commonly known systems, namely, Schr\"{o}dinger, Pauli, and Klein-Gordon particles. We showed that, for instance, the expression for the Bohm quantum potential, the so-called Weizs\"{a}cker correction, the Madelung equations, and the Chen-Sudan Lagrangian density all emerge naturally within our unifying theory as special cases. We also obtained an alternative, manifestly Lagrangian representation of Takabayasi equations for a Pauli particle. Our model yields the dynamics of the spin vector $\vec{S}$ directly from the VP and employs two, rather than three, equations; hence, $S = 1/2$ is ensured irrespective of initial conditions.

Our results can be viewed as a generalization of and a complement to reparameterizations of the FQL that were reported in literature earlier \cite{foot:other}. In addition to this, the new approach has several important applications. First of all, it enables a first principle classical Lagrangian description of most general, relativistic vector particles such as a Dirac electron, which is a long-standing problem (see, \eg \Ref{ref:gaioli98}). Second, the same theory is applicable \textit{as~is} to a GO description of classical vector waves, since the fundamental Lagrangian of classical waves is identical to the FQL \cite{my:wkin}. This allows accounting for polarization effects and mode conversion directly in ray equations. This also allows to extend the recent studies of ponderomotive forces on scalar waves and particles \cite{my:lens} to more general, vector waves and particles. Third, calculations of even purely classical dynamics can be simplified through the application of quantum Lagrangians. This is because, in contrast to classical equations for point particles, semiclassical equations are linear and thus sometimes are easier to work with. These and other applications of the theory presented here will be discussed in follow-up papers, including \Ref{foot:myqdirac}.

The authors thank J.~W. Burby and N.~J. Fisch for valuable discussions. The work was supported by the NNSA SSAA Program through DOE Research Grant No. DE274-FG52-08NA28553, by the U.S. DOE through Contract No. DE-AC02-09CH11466, and by the U.S. DOD NDSEG Fellowship through Contract No. FA9550-11-C-0028.

\appendix

\section{Takabayasi equations}
\label{app:pauli}

In this appendix, we show that the ELEs derived in \Sec{sec:pauli} for a Pauli particle agree with Takabayasi equations \eq{eq:takamom} and \eq{eq:takaspin}.

\subsection{Momentum equation}

To derive the momentum equation, we begin by rewriting the ``spin stress'' introduced by Takabayasi \cite{ref:takabayasi55},
\begin{gather}
\vec{\Pi} \doteq - \frac{1}{m\mc{I}}\, \pd_j \left[\mc{I} (\del S_i) (\pd^j S^i) \right],
\label{eq:Pi}
\end{gather}
in \Eq{eq:takamom}. Specifically, by substituting \Eq{eq:spin_vector} for $\vec{S}$ in \Eq{eq:Pi}, one can write
\begin{gather}
\vec{\Pi} = -\frac{1}{4m\mc{I}} \pd_j \left[ \mc{I} (\del \zeta) (\pd^j \zeta) +\mc{I} \sin^2\zeta ~(\del \vartheta) (\pd^j \vartheta) \right].
\label{eq:spinstress}
\end{gather}
Also, let us use \Eq{eq:dzeta} to express $d_t \vartheta$ and substitute the result in the right hand side of \Eq{eq:momp}. Taking the gradient of the resulting equation yields
\begin{widetext}
\begin{multline}
m d_t \vec{v}= q(\vec{E} + \vec{v} \times \vec{B}) - \del Q 
+ \frac{q}{m}\, \del \left[\vec{S} \cdot \vec{B} 
+ \cot \zeta\, (\pd_\zeta \vec{S}) \cdot \vec{B} \right] 
\\ + \del \left[\frac{\cot \zeta}{4m\mc{I}} \, \del \cdot (\mc{I} \del \zeta) \right]
- \frac{1}{8m}\, \del \left[ (\del \zeta)^2 + (\del \vartheta)^2 \right] .
\label{eq:mv}
\end{multline}
Proceeding with the calculation of the conservation equation for $m\vec{u}$, one obtains
\begin{align}
m \pd_t \vec{u} 
= & - \frac{1}{2}\, \pd_t \left[(\del \vartheta) \cos \zeta \right] \notag \\
= & - \frac{1}{2}\,(\del \vartheta)\, \pd_t \left( \cos \zeta \right) -  \frac{1}{2} \, \cos \zeta \, \del \left(\pd_t \vartheta \right) \notag \\
= & - (\del \vartheta) \left[ -\frac{1}{2} \, (\vec{v} \cdot \del) \cos \zeta + \frac{\cos \zeta}{2 \mc{I}}\, \del \cdot (\mc{I} \vec{u}) 
		+ \frac{1}{4m\mc{I}} \, \del \cdot (\mc{I} \del \vartheta) + \frac{q}{m}\,(\pd_\vartheta \vec{S}) \cdot \vec{B} \right] \notag \\
  & - \cos \zeta \,  \del \left[ - \frac{1}{2}\, (\vec{v}\cdot \del) \vartheta + \frac{1}{4m\mc{I} \sin \zeta} \, \del \cdot (\mc{I} \del \zeta) 
		+ \frac{q}{m \sin \zeta}\, (\pd_\zeta \vec{S}) \cdot \vec{B} \right] \notag \\
= & - m (\vec{v}\cdot \del) \vec{u} -m (\del v_i) u^i + \frac{m\vec{u}}{\mc{I}}\, \del \cdot (\mc{I} \vec{u}) 
		- \frac{(\del \vartheta)}{4m\mc{I}}\, \del \cdot (\mc{I} \del \vartheta) \notag \\
	& - \frac{q}{m} \, [(\pd_\vartheta \vec{S}) \cdot \vec{B}] \, \del \vartheta
	  - \cos \zeta \, \del \left[\frac{q}{m\sin \zeta} \, (\pd_\zeta \vec{S}) \cdot \vec{B} \right]
		- \cos \zeta \, \del \left[\frac{1}{4m\mc{I} \sin \zeta}\, \del \cdot (\mc{I} \del \zeta) \right] \notag \\
= & - m (\vec{v}\cdot \del) \vec{u} - m (\vec{u} \cdot \del) \vec{v} +q \vec{u} \times \vec{B}  
	  + \frac{\cos \zeta}{4m \mc{I}} \, (\del \vartheta) \del \cdot [(\del \vartheta) \mc{I} \cos \zeta]  
    - \frac{(\del \vartheta)}{4m\mc{I}}\, \del \cdot (\mc{I} \del \vartheta) \notag \\
  & - \frac{q}{m}\,[(\pd_\vartheta \vec{S}) \cdot \vec{B}]  (\del \vartheta) 
    - \cos \zeta\, \del \left[\frac{q}{m \sin \zeta}\,(\pd_\zeta \vec{S}) \cdot \vec{B} \right]
		- \cos \zeta \, \del \left[\frac{1}{4m\mc{I} \sin \zeta} \, \del \cdot (\mc{I} \del \zeta) \right],
\label{eq:mu}
\end{align}
where we used \Eqs{eq:dvartheta} and \eq{eq:dzeta}. Let us add \Eqs{eq:mv} and \eq{eq:mu} and also add $(\vec{u} \cdot \del) \vec{u}$ to both sides in order to complete the convective derivative $\hat{\mc{D}} \doteq \pd_t + \vec{V} \cdot \del$. One then gets
\begin{align}
m \hat{\mc{D}} \vec{V}
= &\, q(\vec{E} + \vec{V} \times \vec{B}) - \del Q + \frac{q}{m}\, S^i (\del B_i) + m (\vec{u} \cdot \del) \vec{u}
	  - \frac{(\del \zeta)}{4m\mc{I}} \, \del \cdot (\mc{I} \del \zeta) - \frac{1}{8m} \,\del [(\del \zeta)^2] \notag \\
  & - \frac{\sin^2 \zeta}{4m\mc{I}} \, (\del \vartheta)\, \del \cdot (\mc{I} \del \vartheta) 
    - \frac{\sin^2 \zeta}{4m} \, (\del \vartheta \cdot \del ) (\del \vartheta)  
		- \frac{\sin \zeta \cos \zeta }{2m} \, (\del \vartheta)\,[(\del \vartheta) \cdot (\del \zeta)] \notag \\
= &\, q(\vec{E} + \vec{V} \times \vec{B}) -\del Q + \frac{q}{m} \, S^i (\del B_i) 
		- \frac{1}{4m\mc{I}} \, \pd_j \left[ \mc{I} (\del \zeta) (\pd^j \zeta) +\mc{I} \sin^2\zeta \, (\del \vartheta) (\pd^j \vartheta) \right] \notag \\
= &\,	q(\vec{E} + \vec{V} \times \vec{B}) -\del Q + \frac{q}{m} \, S^i (\del B_i)
		- \frac{1}{m \mc{I}} \, \pd_j \left[ \mc{I} (\del S_i) (\pd^j S^i ) \right],
\end{align}
where we substituted \Eq{eq:spinstress} at the end. Hence, we see that the ELEs derived in \Sec{sec:pauli} lead to \Eq{eq:takamom}.

\subsection{Spin equation}

To derive the spin equation, let us start by re-expressing the ``spin torque'' introduced by Takabayasi \cite{ref:takabayasi55},
\begin{gather}
\vec{M} \doteq \frac{1}{m \mc{I}}\, [\vec{S} \times \pd_j(\mc{I} \, \pd^j \vec{S})],
\end{gather}
in terms of $\vartheta$ and $\zeta$. Specifically, we get
\begin{align}
M_x 
= &\, \frac{1}{m \mc{I}} \, S_y \del \cdot (\mc{I} \del S_z) - \frac{1}{m \mc{I}}\, S_z \del \cdot (\mc{I} \del S_y) \notag \\
= &\, \frac{1}{4m \mc{I}} \, \sin \zeta \sin \vartheta\, \del \cdot [\mc{I} \del(\cos \zeta)] -
    \frac{1}{4m \mc{I}} \, \cos \zeta \, \del \cdot [\mc{I} \del(\sin \zeta \sin \vartheta)] \notag \\
= & - \frac{1}{4m \mc{I}} \, \sin \zeta \sin \vartheta \, \del \cdot [\mc{I} (\del \zeta) \sin \zeta] 
    - \frac{1}{4m \mc{I}} \, \cos \zeta \, \del \cdot [\mc{I} (\del \zeta) \cos \zeta \sin \vartheta]  
    - \frac{1}{4m \mc{I}} \, \cos \zeta \, \del \cdot [\mc{I} (\del \vartheta) \sin \zeta \cos \vartheta] \notag \\
= &\, \frac{\sin \vartheta}{4m}
    \left[
      (\del \vartheta)^2 \sin \zeta \cos \zeta - \frac{1}{\mc{I}} \, \del \cdot (\mc{I} \del \zeta) 
     \right] \notag \\
  & - \frac{1}{4m} \, (\del \zeta)^2 \sin \zeta \cos \zeta \sin \vartheta 
    - \frac{1}{4m} \, \cos \zeta (\del \zeta)  \cdot \del (\cos \zeta \sin \vartheta) 
    - \frac{1}{4m \mc{I}} \, \cos \zeta \cos \vartheta \, \del \cdot [\mc{I} (\del \vartheta) \sin \zeta]  \notag \\
= &\, \frac{\sin \vartheta}{4m} \left[(\del \vartheta)^2 \sin \zeta \cos \zeta - \frac{1}{\mc{I}} \, \del \cdot (\mc{I} \del \zeta) \right]
    - \frac{1}{4m \mc{I}} \, \cot \zeta \cos \vartheta \, \del \cdot [\mc{I} (\del \vartheta) \sin^2 \zeta],\label{eq:Mx}
\end{align}
\begin{align}
M_y 
= &\, \frac{1}{m \mc{I}} \, S_z \del \cdot (\mc{I} \del S_x) - \frac{1}{m \mc{I}} \, S_x \del \cdot (\mc{I} \del S_z) \notag \\
= &\, \frac{1}{4m \mc{I}} \, \cos \zeta \, \del \cdot [\mc{I} \del(\sin \zeta \cos \vartheta)]
    - \frac{1}{4m \mc{I}} \, \sin \zeta \cos \vartheta \, \del \cdot [\mc{I} \del(\cos \zeta)] \notag \\
= &\, \frac{1}{4m \mc{I}} \, \cos \zeta \, \del \cdot [\mc{I} (\del \zeta) \cos \zeta \cos \vartheta]  
    - \frac{1}{4m \mc{I}} \, \cos \zeta \, \del \cdot [\mc{I} (\del \vartheta) \sin \zeta \sin \vartheta] 
    + \frac{1}{4m \mc{I}} \, \sin \zeta \cos \vartheta \, \del \cdot [\mc{I} (\del \zeta) \cos \zeta]  \notag \\
= & - \frac{\cos \vartheta}{4m} \left[(\del \vartheta)^2 \sin \zeta \cos \zeta -  \frac{1}{\mc{I}} \, \del \cdot (\mc{I} \del \zeta) \right]
- \frac{1}{4m \mc{I}} \, \cot \zeta \sin \vartheta \, \del \cdot [\mc{I} (\del \vartheta) \sin^2 \zeta]  ,
    \label{eq:My}
\end{align}
\begin{align}
M_z 
= & \,\frac{1}{m \mc{I}}\, S_x  \del \cdot (\mc{I} \del S_y) - \frac{1}{m \mc{I}}\, S_y \del \cdot (\mc{I} \del S_x) \notag \\
= & \,\frac{1}{4m \mc{I}}\, \sin \zeta \cos \vartheta\, \del \cdot [ \mc{I} \del(\sin \zeta \sin \vartheta)] -
\frac{1}{4m \mc{I}}\, \sin \zeta \sin \vartheta \, \del \cdot [\mc{I} \del(\cos \zeta \cos \vartheta)] \notag \\
= & \,\frac{\sin \zeta \cos \zeta}{2m} \, (\del \zeta) \cdot (\del \vartheta) 
  +   \frac{\sin^2 \zeta}{4m \mc{I}}\,(\del \mc{I}) \cdot (\del \vartheta)\, 
  +   \frac{\sin^2 \zeta}{4m}\, (\del^2 \vartheta)  \notag \\
= & \,\frac{1}{4m \mc{I}} \, \del \cdot [\mc{I} (\del \vartheta) \sin^2 \zeta].\label{eq:Mz}
\end{align}

Hence, the conservation equations for the spin vector $\vec{S}$ can be expressed as follows:
\begin{align}
\hat{\mc{D}} S_x
= & \,\frac{1}{2} \, \pd_t (\sin \zeta \cos \vartheta) + \frac{1}{2}\,(\vec{V} \cdot \del) (\sin \zeta \cos \vartheta)\notag \\
= & \,\frac{1}{2} \, \cos \zeta \cos \vartheta\,\hat{\mc{D}} \zeta
  - \frac{1}{2} \, \sin \zeta \sin \vartheta \, \hat{\mc{D}} \vartheta \notag \\
= & - \cot \zeta \, \cos \vartheta \, \hat{\mc{D}} S_z
  - \frac{1}{2} \, \sin \vartheta \sin \zeta \, \hat{\mc{D}} \vartheta \notag \\
= & - \cot \zeta \cos \vartheta 
      \left\{
      \frac{1}{4m \mc{I}}\,\del \cdot [\mc{I} (\del \vartheta) \sin^2 \zeta] + \frac{q}{m}\,(\pd_\vartheta \vec{S}) \cdot \vec{B} 
      \right\} \notag \\
  & - \sin \vartheta 
      \left[ 
       \frac{1}{2} \, \sin \zeta\, (\vec{u} \cdot \del) \vartheta + \frac{1}{4m \mc{I}} \, \del \cdot (\mc{I} \del \zeta) 
       + \frac{q}{m} \,(\pd_\zeta \vec{S}) \cdot \vec{B}
       \right] \notag \\
= & - \cot \zeta \cos \vartheta 
      \left[
       \frac{1}{4m \mc{I}} \, \del \cdot [\mc{I} (\del \vartheta) \sin^2 \zeta]
       + \frac{q}{m}\, (\pd_\vartheta \vec{S}) \cdot \vec{B} \
      \right] \notag \\
  & - \sin \vartheta 
      \left[
       - \frac{1}{4m} \, (\del \vartheta)^2 \sin \zeta \cos \zeta 
       + \frac{1}{4m \mc{I}} \, \del \cdot (\mc{I} \del \zeta) 
       + \frac{q}{m} \, (\pd_\zeta \vec{S}) \cdot \vec{B}
      \right] \notag \\
= & - \frac{1}{4m \mc{I}} \, \cot \zeta \cos \vartheta \, \del \cdot [\mc{I} (\del \vartheta) \sin^2 \zeta]  
    + \sin \vartheta
      \left[
       \frac{1}{4m} \, (\del \vartheta)^2 \sin \zeta \cos \zeta 
       -  \frac{1}{4m \mc{I}} \, \del \cdot (\mc{I} \del \zeta) 
      \right] + \frac{q}{m} \, (\vec{S}  \times \vec{B})_x \notag\\
= & \,\frac{1}{m \mc{I}} \, [\vec{S} \times \pd_j ( \mc{I} \pd^j \vec{S})]_x + \frac{q}{m} \, (\vec{S}  \times \vec{B})_x,
\label{eq:Sx}
\end{align}
where we substituted \Eq{eq:Mx}; also,
\begin{align}
\hat{\mc{D}} S_y
= & \,\frac{1}{2} \, \pd_t (\sin \zeta \sin \vartheta) + \frac{1}{2} \, (\vec{V} \cdot \del) (\sin \zeta \sin \vartheta)\notag \\
= & \,\frac{1}{2} \, \cos \zeta \sin \vartheta \, \hat{\mc{D}} \zeta
  + \frac{1}{2} \, \sin \zeta \cos \vartheta \, \hat{\mc{D}} \vartheta \notag \\
= & - \cot \zeta \, \sin \vartheta \, \hat{\mc{D}} S_z
  + \frac{1}{2} \, \cos \vartheta \sin \zeta \, \hat{\mc{D}} \vartheta \notag \\
= & - \cot \zeta \sin \vartheta \left\{
     \frac{1}{4m \mc{I}} \, \del \cdot [\mc{I} (\del \vartheta) \sin^2 \zeta] + \frac{q}{m} \, (\pd_\vartheta \vec{S} ) \cdot \vec{B}
    \right\} \notag \\
   & + \cos \vartheta \left[
     \frac{1}{2} \, \sin \zeta \, (\vec{u} \cdot \del) \vartheta +
     \frac{1}{4m \mc{I}} \, \del \cdot (\mc{I} \del \zeta) + 
     \frac{q}{m} \, (\pd_\zeta \vec{S}) \cdot \vec{B} \right] \notag \\
= & - \frac{1}{4m \mc{I}} \, \cot \zeta \sin \vartheta \, \del \cdot [\mc{I} (\del \vartheta) \sin^2 \zeta]   
    - \cos \vartheta \left[
       \frac{1}{4m} \, (\del \vartheta)^2 \sin \zeta \cos \zeta -
       \frac{1}{4m \mc{I}} \, \del \cdot (\mc{I} \del \zeta) 
      \right] + \frac{q}{m} \, (\vec{S}  \times \vec{B})_y \notag \\
= & \,\frac{1}{m \mc{I}} \, [\vec{S} \times \pd_j (\mc{I} \pd^j \vec{S})]_y + \frac{q}{m} \, (\vec{S}  \times \vec{B})_y,
\label{eq:Sy}
\end{align}
where we substituted \Eq{eq:My}. Finally, the equation for $S_z$ is obtained directly from \Eq{eq:dvartheta},
\begin{align}
\mc{I} [\pd_t S_z + (\vec{v} \cdot \del) S_z] + S_z [\pd_t \mc{I} + \del \cdot (\mc{I} \vec{v})] 
= \frac{1}{4m} \, \del \cdot (\mc{I} \del \vartheta) + \frac{q \mc{I}}{m}\, (\pd_\vartheta \vec{S}) \cdot \vec{B}.
\end{align}
Using the ACT, one then gets
\begin{align}
\hat{\mc{D}} S_z = & \frac{1}{\mc{I}} \, [\mc{I} (\vec{u} \cdot \del) S_z + S_z \del \cdot (\mc{I} \vec{u})]
+ \frac{1}{4m \mc{I}} \, \del \cdot (\mc{I} \del \vartheta ) + \frac{q}{m}\, (\pd_\vartheta \vec{S}) \cdot \vec{B}  \notag \\
= & \frac{1}{\mc{I}}\, \del \cdot (S_z \mc{I} \vec{u}) + \frac{1}{4m \mc{I}}\, \del \cdot (\mc{I} \del \vartheta)
 +  \frac{q}{m}\, (\pd_\vartheta \vec{S}) \cdot \vec{B}  \notag \\
= & \frac{1}{4m \mc{I}} \, \del \cdot [\mc{I} (\del \vartheta) \sin^2 \zeta]
+ \frac{q}{m}\,(\vec{S} \times \vec{B})_z\notag \\
= & \frac{1}{m \mc{I}}\, [\vec{S} \times \pd_j (\mc{I}\, \pd^j \vec{S})]_z + \frac{q}{m}\,(\vec{S} \times \vec{B})_z,
\label{eq:Sz}
\end{align}
where we substituted \Eq{eq:Mz}. Considered together as a vector equation, \Eqs{eq:Sx}-\eq{eq:Sz} coincide with \Eq{eq:takaspin}.\\

\end{widetext}


\begin{thebibliography}{10}
\bibitem{foot:adhoc}
For given dynamic equations, a Lagrangian is usually guessed or constructed
  using Lagrange multipliers \cite{ref:seliger68}, neither of which methods is
  universal. The existing methods of deriving classical variational principles
  from quantum mechanics \cite{foot:other} are also limited, as we discuss
  below.
\bibitem{foot:df}
Another commonly used form of the FQL is $\mcc{L} = \psi^\dag (i\pd_t \psi -
  \hat{H}\psi)$ \cite{ref:feldmeier00}. This formula differs from \Eq{eq:gl}
  only by a perfect time derivative and thus is physically equivalent, \ie
  leads to the same ELEs. Here we prefer \Eq{eq:gl}, since the corresponding
  $\mcc{L}$ is real.
\bibitem{foot:pose}
Here we assumed that the particle has positive energy. Otherwise, signs will be
  different, because the scalar product in the $\psi$ space is not positively
  defined \cite{my:wkin}.
\end{thebibliography}
\end{document}